\documentclass[twocolumn,prl,aps,superscriptaddress]{revtex4}
\usepackage{graphicx}
\usepackage{color}
\begin{document}

\title{Coexistence of Half-Metallic Itinerant Ferromagnetism with Local-Moment Antiferromagnetism in Ba$_{0.60}$K$_{0.40}$Mn$_2$As$_2$}

\author {Abhishek Pandey}
\altaffiliation{apandey@ameslab.gov}
\author{B. G. Ueland}\author{S. Yeninas}
\author{A. Kreyssig}\author{A. Sapkota}
\affiliation{Ames Laboratory and Department of Physics and Astronomy, Iowa State University, Ames, Iowa 50011, USA}
\author{Yang Zhao}
\affiliation{NIST Center for Neutron Research, National Institute of Standards and Technology, Gaithersburg, Maryland 20899, USA}
\affiliation{Department of Materials Science and Engineering, University of Maryland, College Park, Maryland 20742, USA}
\author{J.~S.~Helton}
\author{J. W. Lynn}
\affiliation{NIST Center for Neutron Research, National Institute of Standards and Technology, Gaithersburg, Maryland 20899, USA}
\author{R. J. McQueeney}
\author{Y. Furukawa}
\author{A. I. Goldman}
\author{D. C. Johnston}
\altaffiliation{johnston@ameslab.gov}
\affiliation{Ames Laboratory and Department of Physics and Astronomy, Iowa State University, Ames, Iowa 50011, USA}
\date{July 9, 2013}

\begin{abstract}

Magnetization, nuclear magnetic resonance, high-resolution x-ray diffraction and magnetic field-dependent neutron diffraction measurements reveal a novel magnetic ground state of Ba$_{0.60}$K$_{0.40}$Mn$_{2}$As$_{2}$ in which itinerant ferromagnetism (FM) below a Curie temperature $T_{\rm C}\approx 100$~K arising from the doped conduction holes coexists with collinear antiferromagnetism (AFM) of the Mn local moments that order below a N\'eel temperature $T_{\rm N} = 480$~K\@.  The FM~ordered moments are aligned in the tetragonal $ab$~plane and are orthogonal to the AFM~ordered Mn moments that are aligned along the $c$~axis.  The magnitude and nature of the low-$T$ FM ordered moment correspond to complete polarization of the doped-hole spins (half-metallic itinerant FM) as deduced from magnetization and $ab$-plane electrical resistivity measurements.

\end{abstract}


\maketitle

Charge-carrier doping of ordered magnetic states can lead to coupling between magnetism and charge transport phenomena, resulting in exotic properties such as high-temperature superconductivity and half-metallic ferromagnetism (FM). The recent discovery of high-temperature superconductivity in iron pnictides \cite{Kamihara-2008, Takahashi-2008, Rotter-2008, Christianson-2008, Johnston-2010, Paglione-2010} opened up an entirely new playground for exploring insulating or narrow-band metallic behavior since some of them have aspects of both local-moment and itinerant magnetism \cite{Johnston-2010,Paglione-2010,Dai-2012, Si-2008, Qazilbash-2009}.  An important related material is ${\rm BaMn_2As_2}$ which has the same tetragonal ${\rm ThCr_2Si_2}$-type crystal structure (space group $I4/mmm$) as the high-temperature superconductor parent compound ${\rm BaFe_2As_2}$ \cite{Rotter-2008}, where the Mn and Fe atoms, respectively, occupy the corners of a planar square lattice.  The itinerant spin-density-wave antiferromagnetic (AFM) ground state of ${\rm BaFe_2As_2}$ has an ordering temperature $T_{\rm N}=137$~K and ordered magnetic moment at low~$T$ of $\mu \sim 1~\mu_{\rm B}$/Fe \cite{Johnston-2010, Kitagawa-2008} where $\mu_{\rm B}$ is the Bohr magneton.  In contrast, ${\rm BaMn_2As_2}$ has an insulating ground state with collinear N\'eel-type (G-type) local-moment AFM order below $T_{\rm N}$ = 625~K in which each Mn$^{+2}$ ion with spin~$S=5/2$ and $\mu=3.9~\mu_{\rm B}$/Mn is oppositely directed to the moment of each of its nearest-neighbors \cite{Johnston-2010, Singh-2009a, An-2009, Singh-2009b, Johnston-2011}. This ground state is similar to those of the AFM insulator parent compounds (e.g., ${\rm La_2CuO_4}$ \cite{Johnston-1997, Kastner-1998}) of layered cuprate high-temperature superconductors \cite{Bednorz-1986}, where the Cu$^{+2}$ cations with $S = 1/2$ also occupy the corners of a square lattice and exhibit G-type AFM order.

${\rm BaMn_2As_2}$ can be made metallic by substituting K for Ba to form hole-doped Ba$_{1-x}$K$_x$Mn$_2$As$_2$ \cite{Pandey-2012,Bao-2012} or by applying pressure \cite{Satya-2011}. The AFM ordering in Ba$_{1-x}$K$_x$Mn$_2$As$_2$ is very robust.  In particular, the $\mu$ is nearly constant at $\mu \approx 4.0~\mu_{\rm B}$/Mn as the doping level changes from the insulating composition $x=0$ to metallic compositions from $x=0.016$ to $x=0.40$ that we have studied, and with the ordered moment oriented along the $c$~axis throughout the doping series \cite{Singh-2009b,Pandey-2012,Lamsal-2013}.  Furthermore, the $T_{\rm N}$ associated with the Mn local moments decreases by only 20\% from 625(1)~K for $x = 0$ to 480(2)~K for $x = 0.40$.  These results suggest weak coupling between the itinerant doped holes and the Mn spins.

Here we report the discovery of a novel magnetic structure in single crystals of Ba$_{0.60}$K$_{0.40}$Mn$_{2}$As$_{2}$ probed using magnetization $M$, magnetic susceptibility $\chi\equiv M/H$ where $H$ is the applied magnetic field, $H$-dependent neutron and high-resolution x-ray diffraction, nuclear magnetic resonance (NMR) and $ab$-plane electrical resistivity $\rho_{ab}$ measurements. Our measurements consistently demonstrate a coexistence below $\approx 100$~K of itinerant FM of the doped holes, with an ordering temperature $T_{\rm C}\approx 100$~K, with collinear Mn local-moment AFM ($T_{\rm N} =480$~K), where the ordered moments in the two magnetic structures are aligned \emph{perpendicular} to each other.  Furthermore, we infer that the FM at $T\to0$ arises from complete spin polarization of the doped holes, called half-metallic FM\@.  This FM component was not present in the previously studied samples of Ba$_{1-x}$K$_x$Mn$_2$As$_2$ with K concentrations of 1.6 and 5\% \cite{Pandey-2012}, and is qualitatively different from the half-metallic FM in, e.g., La$_{1-x}$Ca$_{x}$MnO$_3$ \cite{Tokura-2006}, where the conduction electron moments are polarized by strong Hund coupling to the FM-ordered local Mn moments and hence align \emph{parallel} to them. 

\begin{figure}
\includegraphics[width=3.3in]{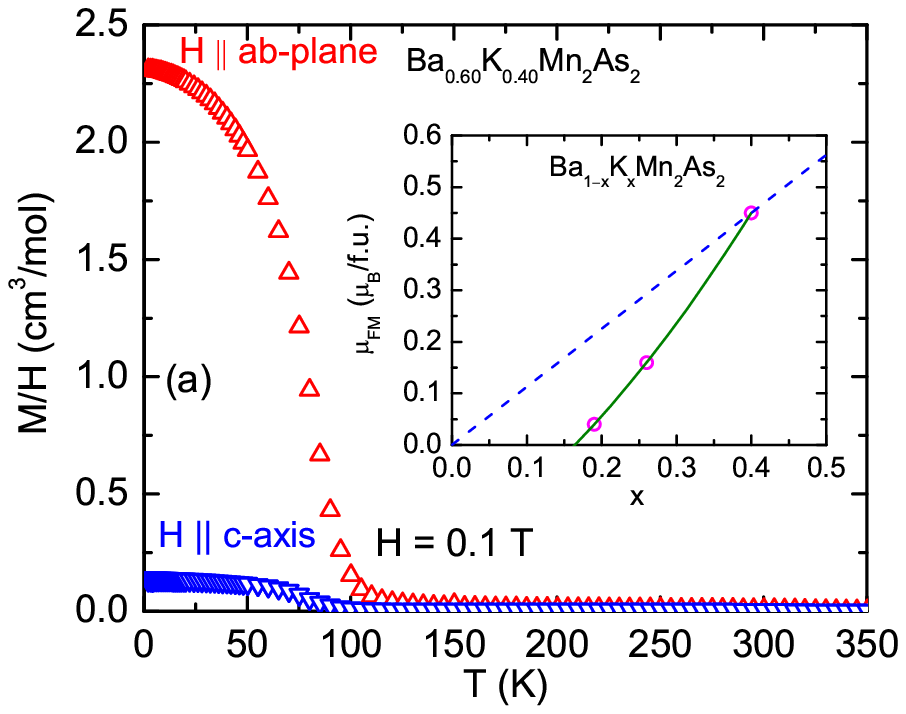}\vspace{0.05in}
\includegraphics[width=3.2in]{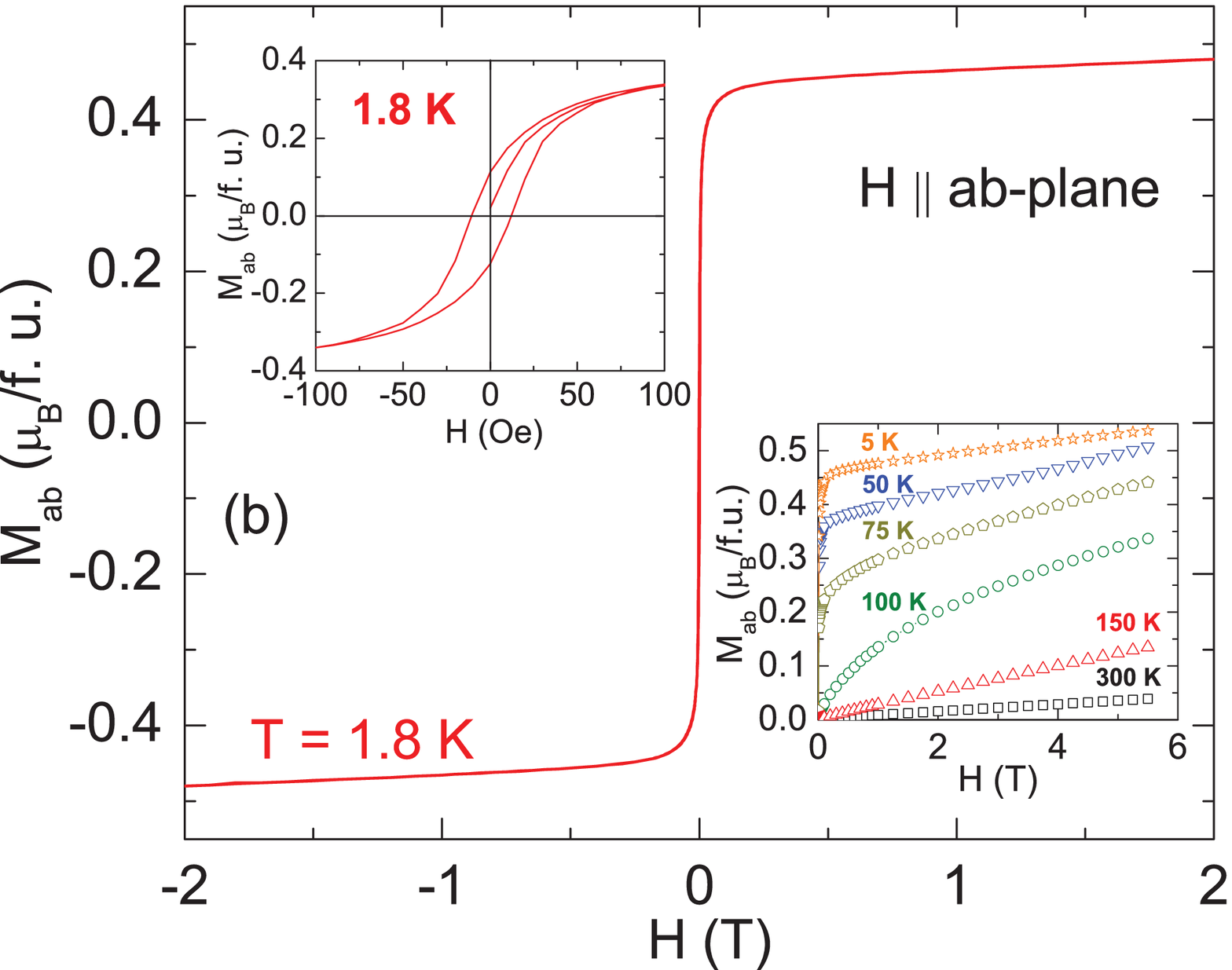}
\caption{(a) $\chi \equiv M/H$ versus $T$ of Ba$_{0.60}$K$_{0.40}$Mn$_2$As$_2$. Zero-field-cooled  and field-cooled data are indistinguishable. Inset: FM ordered moment $\mu_{\rm FM}$ at low~$T$ versus $x$ in Ba$_{1-x}$K$_x$Mn$_2$As$_2$.  The data for $x=0.19$ and~0.26 are from \cite{Bao-2012}.  The solid green curve is a guide to the eye.  The blue dashed line is the expected behavior for full doped-hole spin polarization with $\mu_{\rm FM} = xgS\mu_{\rm B}$/f.u., $g=2.25$ and $S=1/2$ which fit the datum for $x=0.40$.  (b)~Magnetization $M_{ab}$ versus $H$ at 1.8~K with ${\bf H} \parallel ab$~plane. Upper inset: Expanded plot of $M_{ab}(H)$ at 1.8~K\@.  Lower inset: $M_{ab}(H)$ isotherms between 5 and 300~K, inclusive.}
\label{fig:Figure_K_MTMH}
\end{figure}

Single crystals of ${\rm Ba_{0.60}K_{0.40}Mn_2As_2}$ were grown using the self-flux solution growth technique \cite{Lamsal-2013}. The composition of the crystals was determined by wavelength-dispersive x-ray spectroscopy.  Four-probe $\rho_{ab}(T)$ and $M(H,T)$ measurements were carried out using Quantum Design, Inc., instruments.  Neutron diffraction measurements were performed on a single crystal using the BT-7 Double Focusing Triple Axis Spectrometer at the NIST Center for Neutron Research \cite{Lynn-2012}. The sample was aligned in the $(h\ 0\ \ell)$ scattering plane, and a vertical-field superconducting magnet was used to apply $H \leq 1$~T along the [0 1 0] direction (out of the scattering plane). The magnet was cooled from high~$T$ prior to the experiments to ensure that no remnant field was present.  Neutrons with incident and final wavelengths $\lambda = 2.359$~\AA\ were selected using the (0~0~2) reflections of a pyrolytic graphite monochromator and analyzer.  NMR measurements were carried out on $^{75}$As nuclei (nuclear spin $I = 3/2$) using a homemade phase-coherent spin-echo pulse spectrometer. 

From the robust nature of the Mn local-moment AFM in Ba$_{1-x}$K$_x{\rm Mn_2As_2}$ described above, the anisotropic $\chi(T)$ of the Mn sublattice in ${\rm Ba_{0.60}K_{0.40}Mn_2As_2}$ is expected to be similar to that of undoped ${\rm BaMn_2As_2}$ \cite{Singh-2009a, Johnston-2011}.  However, as shown in Fig.~\ref{fig:Figure_K_MTMH}(a), the $\chi_{ab}$ with ${\bf H} \parallel ab$~plane instead exhibits a huge enhancement on cooling below 100~K, whereas $\chi_{c}$ with ${\bf H} \parallel c$~axis shows an increase, but much smaller, indicating the occurrence of a FM transition below a Curie temperature $T_{\rm C}\approx 100$~K with the FM ordered moments oriented in the $ab$~plane.  The $\chi_{ab}$ and $\chi_{c}$ values at 2~K are three and two orders of magnitude larger than the respective values for the parent compound ${\rm BaMn_2As_2}$ \cite{Singh-2009a,Johnston-2011}.  This FM coexists with the G-type AFM ordering of the Mn spins below $T_{\rm N} = 480$~K \cite{Lamsal-2013}.  We attribute the $\chi_c(T\to0)$ value to a $\approx 4^\circ$ misalignment of {\bf H} with the $c$~axis of the crystal.

The FM is confirmed in Fig.~\ref{fig:Figure_K_MTMH}(b) from $M_{ab}(H)$ measurements at 1.8~K\@.  The data show a very nonlinear behavior where $M_{ab}$ saturates to a nearly constant value at fields above about 100~Oe.  Extrapolating the high-field linear behavior to $H=0$ gives $\mu_{\rm FM}(T = 1.8~{\rm K}) = 0.45(1)~\mu_{\rm B}$/f.u.\ (f.u.\ denotes formula unit).  The expanded plot in the upper inset shows that the FM is extremely soft since the saturation is nearly complete at a very small field of $\sim 100$~Oe  and the coercive field is only about 10~Oe. $M_{ab}(H)$ isotherms obtained both above and below 100~K are shown in the lower inset of Fig.~\ref{fig:Figure_K_MTMH}(b).  The spontaneous magnetization is clearly visible in the isotherms at 75~K and below.

One explanation for the FM below $T_{\rm C}$ in ${\rm Ba_{0.60}K_{0.40}Mn_2As_2}$ with the ordered moment lying in the $ab$~plane is that it originates from canting (tilting) of the already-established Mn ordered moments  ($T_{\rm N} = 480$~K) along the $c$~axis towards the $ab$~plane.  The AFM ordered moment along the $c$~axis below 100~K is $\approx 3.9~\mu_{\rm B}$/Mn~\cite{Lamsal-2013} with two Mn atoms per f.u., so a canting of the Mn moments away from the $c$~axis by an angle of only $3.4^\circ$ could explain the observed in-plane FM moment at low~$T$ of $0.45~\mu_{\rm B}$/f.u.  In the following we use two approaches to rule out this explanation.  We conclude instead that the FM is itinerant, arising from spin polarization of the itinerant doped holes.

\begin{figure}
\includegraphics[width=3.3in]{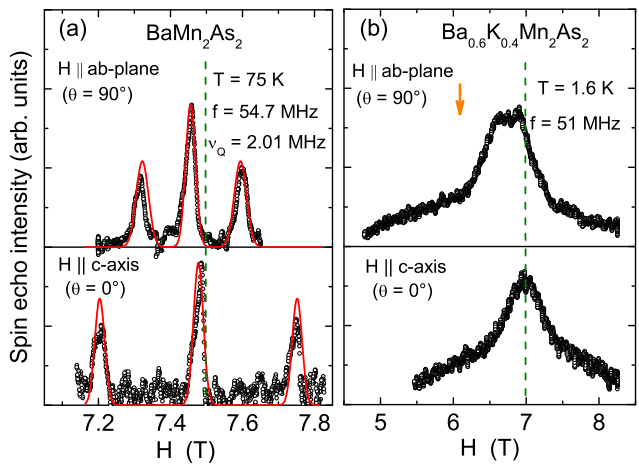}
\caption{$^{75}$As-NMR spectra for single crystals of (a)~BaMn$_2$As$_2$ at $T = 75$~K and (b) Ba$_{0.60}$K$_{0.40}$Mn$_2$As$_2$ at $T = 1.6$~K\@.  The green vertical dashed lines are the zero-shift positions.  The solid red curves in (a) are simulated spectra and the vertical orange arrow in the top part of (b) indicates the expected NMR line position if the FM component of the ordering were caused by canting of the Mn spins.}
\label{fig:Figure_K_NMR}
\end{figure}

The first approach utilizes NMR measurements.  Figure~\ref{fig:Figure_K_NMR}(a) shows field-swept $^{75}$As-NMR spectra in the AFM state of an undoped $\rm{BaMn_2As_2}$ crystal at $T = 75~{\rm K}\ll T_{\rm N} = 625$~K with $H\sim7$--8~T\@. The $^{75}$As site symmetry is such that there is no net contribution to the internal field $H_{\rm int}$ at the $^{75}$As sites due to the G-type AFM ordering of the Mn moments as seen for ${\bf H}\parallel c$~axis in the bottom part of Fig.~\ref{fig:Figure_K_NMR}(a).  We fitted the spectra using the nuclear spin Hamiltonian ${\cal H} = \frac{\gamma}{2\pi}h\vec{I}\cdot\vec{H}_{\rm eff} + \frac{h\nu_{Q}}{6}[3I^{2}_z - I(I + 1)]$, where $h$ is Planck's constant, $H_{\rm eff} = H + H_{\rm int}$ is the effective field at the $^{75}$As site and $\nu_Q$ is the nuclear quadrupole frequency. The spectrum is well reproduced by  $\nu_{Q} = 2.01$~MHz, $H_{\rm int}= 0.013$~T for ${\bf H} \parallel c$~axis and $H_{\rm int} = 0.045$~T for ${\bf H} \parallel ab$~plane at $T = 75$~K [solid red curves in Fig.~\ref{fig:Figure_K_NMR}(a)], and almost independent of~$T$ for $T=4.2$--300~K\@.  The small shift in the spectrum for ${\bf H}\parallel ab$~plane from $H_{\rm int}$ is due to field-induced tilting of the $c$-axis Mn moments towards the $ab$~plane.  From a $K$-$\chi$ analysis for ${\bf H}\parallel ab$~plane, we obtained the hyperfine coupling constant of the $^{75}$As nuclei to a Mn ordered moment as $A_{ab}= 0.99~{\rm T}/\mu_{\rm B}$.  This value is comparable to $A_{ab} = 0.66$~T/$\mu_{\rm B}$ reported for ${\rm BaFe_2As_2}$~\cite{Kitagawa-2008}.  

For $\rm {Ba_{0.60}K_{0.40}Mn_2As_2}$, the $^{75}$As-NMR spectrum at 1.6~K becomes broad and no clear quadrupole splitting is observed as shown in Fig.~\ref{fig:Figure_K_NMR}(b).  We estimate the contribution to the average  internal field at the $^{75}$As site due to the Mn ordered moments in $H\approx 7$~T as $H_{\rm int} \approx 0$ and 0.25~T for ${\bf H} \parallel c$~axis and ${\bf H} \parallel ab$~plane at $T = 1.6$~K, respectively.  If one assumes that the FM saturation moment of about $0.23~\mu_{\rm B}$/Mn in the $ab$~plane at 1.6~K  arises from canting of the Mn ordered moments and utilizes the above value of $A_{ab}$, the $H_{\rm int}$ at the $^{75}$As site is expected to be 0.91~T, and the expected position of the peak is indicated by the vertical arrow in the top panel of Fig~\ref{fig:Figure_K_NMR}(b). This estimated $H_{\rm int}$ value is nearly four times larger than the observed value, suggesting that the FM in the $ab$~plane of ${\rm Ba_{0.60}K_{0.40}Mn_2As_2}$ is not due to canting of the Mn spins.

Our second and more definitive approach to rule out Mn spin canting as the origin of the in-plane FM uses a combination of high-resolution x-ray diffraction and field-dependent magnetic neutron diffraction measurements.  Canting of an otherwise collinear AFM structure typically arises from antisymmetric exchange coupling between the moments via the Dzyaloshinskii-Moriya interaction \cite{Dzyaloshinsky-1958, Moriya-1960b} which for our compound requires a lowering of the crystal symmetry to eliminate the inversion center between the Mn moments.  However, our high-resolution x-ray diffraction  scans along the [1~1~0] and [1~0~0] directions through the (2~2~10) and (3~0~11) Bragg peaks, respectively, in the $T$ range 10--300~K show no change in the peak shape, or splitting, that would signal an orthorhombic distortion of the tetragonal basal plane. From the half-width of the Bragg peaks, we estimate an upper limit for the orthorhombicity as $(a-b)/(a+b) <2.5 \times 10^{-4}$, where $a-b$ is the potential difference in the basal-plane lattice parameters. 

\begin{figure}
\includegraphics[width=3in]{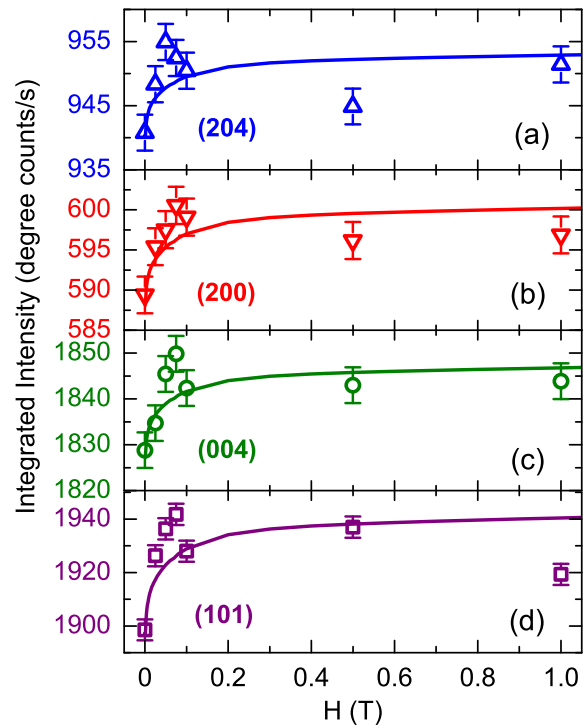}
\caption{Integrated intensity versus $H$ for (a)~(2~0~4), (b)~(2 0 0), (c)~(0 0 4), and (d)~(1 0 1) Bragg peaks at $T = 5$~K\@. Uncertainties are statistical in origin and represent one standard deviation.  The curves are fits by $M_{ab}^2(H)$ in Fig.~\ref{fig:Figure_K_MTMH}(b). }	
\label{fig:Figure_Neutron}
\end{figure}

We carried out neutron diffraction measurements on four Bragg peaks at $T=5$~K with {\bf H} applied in the [0~1~0] direction as shown in Fig.~\ref{fig:Figure_Neutron}.  The (2~0~4), (2~0~0) and (0~0~4) peaks are nuclear Bragg peaks whereas the (1~0~1) Bragg peak also contains a contribution from the G-type AFM structure of the Mn moments below $T_{\rm N} = 480$~K \cite{Lamsal-2013}. For each peak, the integrated intensity $I$ shows an overall increase with increasing $H$ with a sharp onset to a saturation plateau at $H \leq 0.025$~T in agreement with the bulk $M_{ab}(H)$ data in Fig.~\ref{fig:Figure_K_MTMH}(b), as shown by the fits of the $I(H)$ data by $I(H) = C_0 + C_1[M_{ab}(H)]^2$ (solid curves) where $C_0$ and $C_1$ are fitted constants.  This agreement demonstrates that the FM is a bulk effect.  Furthermore, the observed increase in $I(H)$ for {\it all four} Bragg peaks indicates that the FM does not arise from the Mn local moments, as follows.  Since the Mn atoms in the $4d$ sites of the $I4/mmm$ space group only contribute to the $(h\ k\ \ell)$ nuclear and FM reflections with $h+k$ even and $\ell$ even, a FM component from the Mn spins would not contribute to the (101) peak.  In particular, an increase in $I$ with $H$ due to an $H$-dependent FM component at the Mn positions would only occur for the (204), (200) and (004) peaks. In view of the similarity of $I(H)$ of all four Bragg peaks we infer that the enhancements of $I$ for all four peaks with increasing~$H$ do not arise from the Mn spins.  Using the Mn magnetic form factor, the estimated FM ordered moment would be on the order of $0.6~\mu_{\rm B}$/f.u., consistent with the $M_{ab}(H)$ data in Fig.~\ref{fig:Figure_K_MTMH}(b); however, due to the itinerant nature of the FM the spatial distribution of the moment is unknown and hence this estimate is not definitive.

The above experiments conclusively demonstrate that the FM ordered moment aligned in the $ab$~plane of $\rm {Ba_{0.60}K_{0.40}Mn_2As_2}$ is not due to canting of the Mn ordered moments away from the $c$~axis.  The substitution of an average of 0.40 K~atoms for each Ba~atom donates 0.40 conduction holes per f.u.\ to insulating ${\rm BaMn_2As_2}$. Each hole is expected to have a spectroscopic splitting factor $g\approx 2$ and $S=1/2$. Thus if all the doped holes order ferromagnetically, the ordered moment at $T=0$ is expected to be $\mu_{\rm FM} = 0.40gS\mu_{\rm B}/{\rm f.u.}\approx 0.4\,\mu_{\rm B}$/f.u., which is close to the value of $0.45(1)\,\mu_{\rm B}$/f.u.\ determined from our $M_{ab}(H)$ data at 1.8~K in Fig.~\ref{fig:Figure_K_MTMH}(a).

Support for the itinerant hole origin of the FM was obtained from measurements of a co-doped single crystal of ${\rm Ba_{0.61}K_{0.39}(Mn_{0.81}Fe_{0.19})_2As_2}$ with a net charge doping close to zero \cite{SupplInfo}, which we found to be an AFM insulator with properties similar to those of undoped ${\rm BaMn_2As_2}$ \cite{Singh-2009a, An-2009}.  Even though the co-doped compound contains 0.4~K/f.u.\ as in $\rm {Ba_{0.60}K_{0.40}Mn_2As_2}$, no FM is observed in this insulating analogue.

We therefore further examine the possibility that the FM is caused by complete spin polarization (at $T=0$) of the itinerant doped holes, a type of material called a half-metallic FM \cite{Groot-1983,Katsnelson-2008,Park-1998,Muller-2009}.  We first note that the extremely small coercive field of $\sim 10$~Oe and the related very fast approach of the magnetization to saturation with increasing field in Fig.~\ref{fig:Figure_K_MTMH}(b) are similar to the behaviors observed for other half-metallic FMs \cite{Coey-1998, Ritchie-2003}, an $M(H)$ behavior very different from that of local-moment FMs where the anisotropy fields are typically much larger.  This also explains the previous observation of FM in  Ba$_{1-x}$K$_x$Mn$_2$As$_2$ crystals with $x = 0.19$ and 0.26, however with saturation moments of only 0.04 and $0.154~\mu_{\rm B}$/f.u., respectively \cite{Bao-2012}.  Considering these results and ours together suggests that there is a critical hole-doping composition $x_{\rm cr}\approx 0.16$ at which the FM first appears, and above which the $T=0$ fully hole-spin-polarized FM state develops at much higher doping concentrations as illustrated in the inset of Fig.~\ref{fig:Figure_K_MTMH}(a).

\begin{figure}
\includegraphics[width=3.3in]{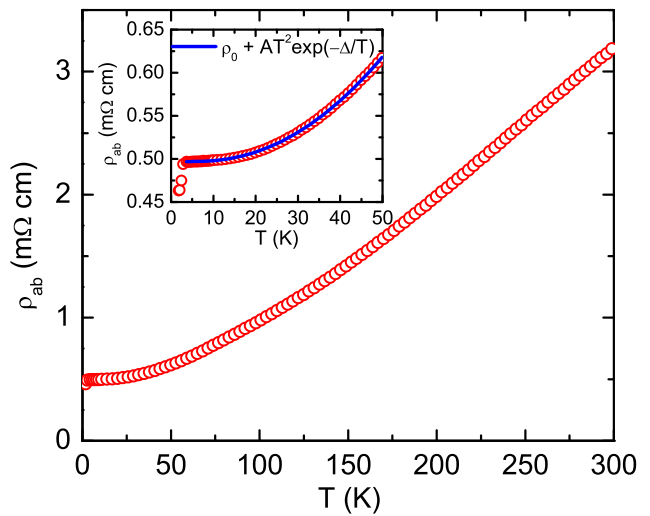}
\caption{In-plane resistivity $\rho_{ab}$ versus $T$ for Ba$_{0.60}$K$_{0.40}$Mn$_2$As$_2$.  Inset: $\rho_{ab}(T)$ from 1.8 to 50~K~\cite{rhoDrop}.  A fit by the given equation (see text) is shown from 3.4 to 50~K with fitting parameters $\rho_0 = 0.50~{\rm m}\Omega\,{\rm cm}$, $A = 7.11(3)\times10^{-5}~{\rm m}\Omega\,{\rm cm}/{\rm K}^2$ and $\Delta = 19.0(2)$~K.}
\label{fig:Figure_Res}
\end{figure}

The $T$ dependence of $\rho$ can be used as a diagnostic for half-metallic FM \cite{Bombor-2013}. The in-plane $\rho_{ab}$ of ${\rm Ba_{0.60}K_{0.40}Mn_2As_2}$ exhibits metallic behavior (Fig.~\ref{fig:Figure_Res}) similar to other Ba$_{1-x}$K$_x$Mn$_2$As$_2$ compositions with $x \geq 0.016$ \cite{Pandey-2012, Bao-2012}.  A single-magnon scattering process leading to a $T^2$ dependence of $\rho$ at low $T$ involves spin-flip transitions and is therefore absent in half-metallic FMs at $T=0$ due to the completely polarized character of the conduction band.  Instead, half-metallic FMs are expected to follow $\rho(T) = \rho_{0} + AT^2\exp(-\Delta/T)$ at low $T$ due to thermally-activated carrier scattering between the spin-split bands, where $\Delta$ is the energy gap in temperature units between the Fermi energy of the majority spins and the bottom of the minority spin band that is empty at $T=0$ \cite{Bombor-2013}. An excellent fit of $\rho_{ab}(T)$ of ${\rm Ba_{0.60}K_{0.40}Mn_2As_2}$ in Fig.~\ref{fig:Figure_Res} by this expression, but not by a $T^2$ dependence, is obtained from 3.4 to 50~K as shown in the inset of Fig.~\ref{fig:Figure_Res} \cite{rhoDrop}.  This agreement strongly supports our conclusion about the half-metallic character of the FM in ${\rm Ba_{0.60}K_{0.40}Mn_2As_2}$.

In summary, our magnetization and neutron diffraction measurements of ${\rm Ba_{0.60}K_{0.40}Mn_2As_2}$ crystals revealed bulk FM ordering below $T_{\rm C}\approx 100$~K with an ordered moment $\mu_{\rm FM}(T\to0) = 0.45(1)~\mu_{\rm B}$/f.u.\ aligned in the $ab$-plane.  This FM ordering coexists with G-type AFM ordering of the Mn local moments with $\mu(T\to0)\approx 3.9~\mu_{\rm B}$/Mn and $T_{\rm N} = 480$~K, where the ordered moments are instead aligned along the $c$-axis.  Our NMR, high-resolution x-ray and $H$-dependent neutron diffraction measurements consistently demonstrate that the FM in the $ab$-plane does not arise from canting of the Mn ordered moments away from the $c$-axis.  From the above measurements and additional $\rho_{ab}(T)$ measurements we conclude that ${\rm Ba_{0.60}K_{0.40}Mn_2As_2}$ exhibits a novel magnetic structure in which half-metallic FM coexists  with local moment AFM of the Mn spin lattice, where the ordered moments of these two magnetic substructures are perpendicular to each other.  Important open questions are the microscopic mechanisms of the half-metallic FM and of the perpendicular alignment of the FM and AFM ordered moments.

\acknowledgments

The authors wish to thank J. Lamsal, G. S. Tucker, W. Jayasekara, M. G. Kim, T. W. Heitmann and W. E. Straszheim for their assistance in the early stages of this project. We also thank V. P. Antropov, B. N. Harmon and A. Kaminski for useful discussions.  The research at Ames Laboratory was supported by the U.S. Department of Energy, Office of Basic Energy Sciences, Division of Materials Sciences and Engineering. Ames Laboratory is operated for the U.S. Department of Energy by Iowa State University under Contract No.~DE-AC02-07CH11358.

\clearpage

\section{Supplemental Material: Electrical Resistivity, Magnetization and Magnetic Susceptibility Measurements on ${\rm\bf Ba_{0.61}K_{0.39}(Mn_{0.81}Fe_{0.19})_2As_2}$ Single Crystals}

We demonstrated in the main text that the ferromagnetism in Ba$_{0.60}$K$_{0.40}$Mn$_2$As$_2$ cannot arise from the Mn magnetic moments.  The only alternative is then that it must originate from the doped conduction holes.  This suggests that if Ba$_{0.60}$K$_{0.40}$Mn$_2$As$_2$ could somehow be made insulating, then no ferromagnetism should occur.  In order to test this hypothesis, we grew single crystals of the co-doped compound ${\rm Ba_{0.61}K_{0.39}(Mn_{0.81}Fe_{0.19})_2As_2}$ in which the 0.4 doped holes/f.u.\ arising from the K-doping is closely compensated by 0.4~doped electrons/f.u.\ by replacing 0.2 Mn by Fe which has one more $d$-electron than Mn.  Thus the net charge doping is close to zero.

\begin{figure}[t]
\includegraphics[width=3.3in]{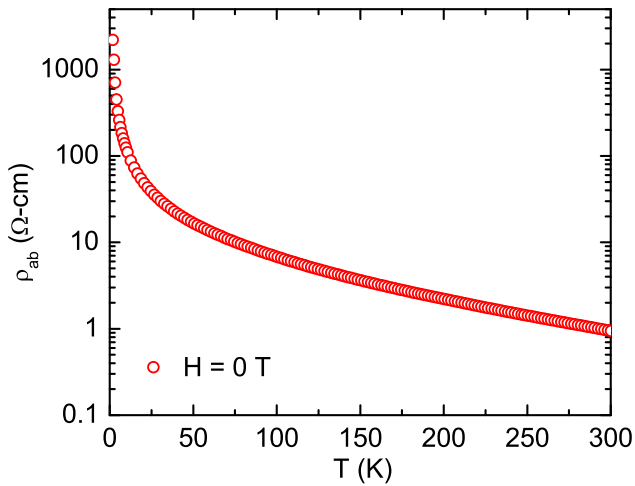}	
\caption{Resistivity $\rho_{ab}$ plotted on a logarithmic scale versus temperature $T$ for co-doped ${\rm Ba_{0.61}K_{0.39}(Mn_{0.81}Fe_{0.19})_2As_2}$.  The data clearly show that ${\rm Ba_{0.61}K_{0.39}(Mn_{0.81}Fe_{0.19})_2As_2}$ has an insulating ground state.}
\label{fig:Supplementary_Fig1}
\end{figure}

\begin{figure}[t]
\includegraphics[width=3.3in]{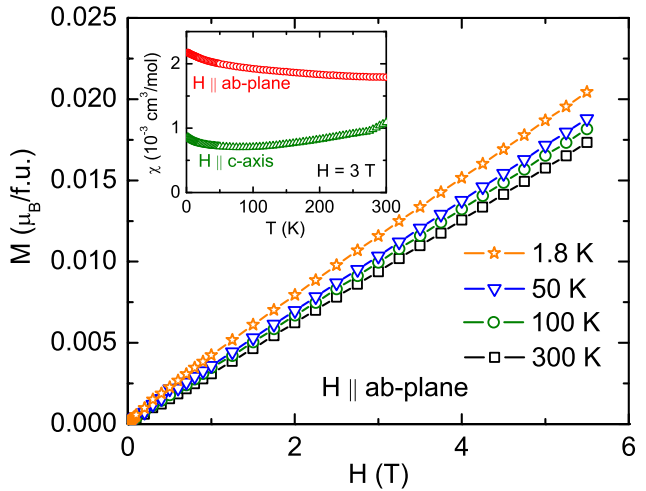}	
\caption{Isothermal magnetization $M$ versus applied magnetic field ${\bf H}\parallel ab$-plane of a crystal of co-doped ${\rm Ba_{0.61}K_{0.39}(Mn_{0.81}Fe_{0.19})_2As_2}$.  The inset shows the magnetic susceptibility $\chi \equiv M/H$ versus temperature $T$ measured at $H=3$~T along two different crystallographic directions. These data show no evidence for ferromagnetism that is clearly seen in single crystals of Ba$_{0.60}$K$_{0.40}$Mn$_2$As$_2$ as described in the main text.}
\label{fig:Supplementary_Fig2}
\end{figure}

Strikingly, the temperature dependence of the in-plane resistivity $\rho_{ab}(T)$ (Fig.~\ref{fig:Supplementary_Fig1}) shows that ${\rm Ba_{0.61}K_{0.39}(Mn_{0.81}Fe_{0.19})_2As_2}$ has an insulating ground state.  Thus the compensated charge doping acts as if there is no doping at all.  

Magnetization versus field $M(H)$ isotherms at four temperatures between 1.8 and~300~K are shown in Fig.~\ref{fig:Supplementary_Fig2}, and the anisotropic magnetic susceptibilities $\chi\equiv M/H$ at $H=3$~T are plotted in the inset of Fig.~\ref{fig:Supplementary_Fig2}.  These data show no evidence for ferromagnetism and instead indicate that  ${\rm Ba_{0.61}K_{0.39}(Mn_{0.81}Fe_{0.19})_2As_2}$ is an antiferromagnet with a N\'eel temperature significantly above room temperature. 

\newpage

To summarize, the above $\rho(T)$, $M(H)$ and $\chi(T)$ behaviors of ${\rm Ba_{0.61}K_{0.39}(Mn_{0.81}Fe_{0.19})_2As_2}$ are similar to those of the undoped antiferromagnetic insulator parent compound ${\rm BaMn_2As_2}$ \cite{Singh-2009a, An-2009}.  In particular, there is no evidence for ferromagnetism at $T\geq 1.8$~K, even though the potassium concentration is the same as in Ba$_{0.60}$K$_{0.40}$Mn$_2$As$_2$ which exhibits in-plane ferromagnetic ordering below 100~K\@.  As stated in the main text, this result provides support for our conclusion that the ferromagnetism in Ba$_{0.60}$K$_{0.40}$Mn$_2$As$_2$ arises from the itinerant doped holes provided by substituting potassium for barium.

\end{document}